\documentclass[aps,prl,twocolumn,superscriptaddress]{revtex4}
\usepackage{amssymb}
\usepackage{amstext}
\usepackage{amscd}
\usepackage{amsthm}
\usepackage{amsopn}
\usepackage{amsmath}
\usepackage{amsbsy}
\usepackage{eufrak}
\usepackage{indentfirst}
\usepackage{graphicx}
\usepackage{epsfig}
\usepackage{bm}
\begin{document}
\title{Erratum: Nonequilibrium Transport in Quantum Impurity Models (Bethe-Ansatz for open systems).
[Phys. Rev. Lett. 96, 216802 (2006)]}
\author{P. Mehta, S-P Chao, N. Andrei}
\maketitle

 In a recent Letter \cite{MA} two of the authors have proposed a new approach
 to quantum impurities out of equilibrium. The approach is based on the construction of scattering
 eigenstates of the Hamiltonian on the infinite line (open system) with the bias appearing as an asymptotic
 boundary condition.  The method was applied to the
 Interacting Resonance Level Model (IRLM)  describing a local level
 $d$
 level coupled to two leads held at different potentials $\mu_i, i =1,2$. Also included in the Hamiltonian
 is an interaction between
  the leads and the level:

\begin{eqnarray}\nonumber
H_{IRLM}= -i \sum_{i=1,2}\int dx \psi_i^{\dagger}(x)\partial\psi_i(x) +\epsilon_d d^{\dagger}d \\
+t(\sum_i \psi_i^{\dagger}(0)d+h.c.)+U \sum_i \psi_i^{\dagger}(0)\psi_{i}(0)d^{\dagger}d
\end{eqnarray}

 However a sign error entered in the equations determining the non-equilibrium Bethe-
momentum densities $\rho_i, i=1,2$. With the sign error the
equations in the Letter correspond to the model with $U$ negative.
When corrected, the equations take the form:
\begin{eqnarray}\nonumber
\rho_1(p)&=& \frac{1}{2\pi}\theta( k^1_o-p)  - \sum_{j=1,2}
\int_{-\infty}^{k^j_o}  {\cal K}(p,k)  \rho_j(k)\;dk \\
 \rho_2(p)&=&
\frac{1}{2\pi}\theta( k^2_o-p) -\sum_{j=1,2} \int_{-\infty}^{k^j_o}
{\cal K}(p,k)  \rho_j(k)\;dk  \; \; \; \; \; \label{BAeq}
\end{eqnarray}

with ${\; \cal K}(p,k)\equiv \frac{U}{\pi} (\epsilon_d-k)/
[(p+k-2\epsilon_d)^2+\frac{U^2}{4}(p-k)^2]$.\\

 The Bethe momenta $p$ in lead $i$  are filled from the lower cut-off $(-D)$ up to $k^i_o$ determined
 from,

\begin{eqnarray}
\int_{-D}^{\mu_i}\frac{1}{2\pi}\;dp=\int_{-D}^{k^i_o}\rho_i(p)\;dp
\end{eqnarray}
and the voltage is $V=\mu_1-\mu_2$.

\begin{figure}[t]
\includegraphics[width=0.85\columnwidth, clip]{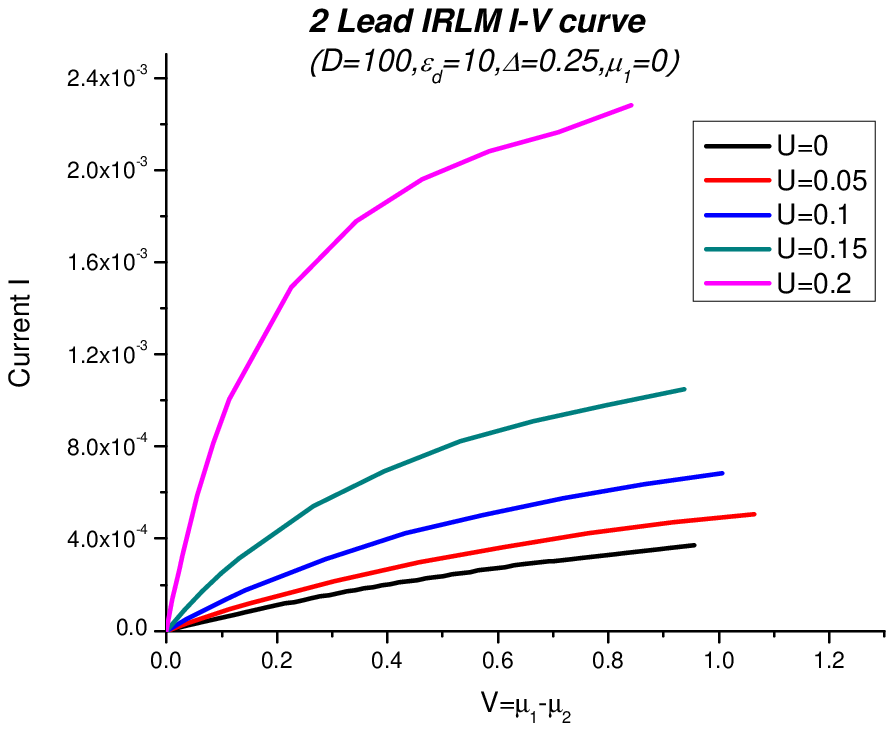}
\caption{ The current as a function of voltage for various $U$ with
$\mu_1=0$ and  $\mu_2$ being lowered. The bandwidth $D=100$,
 the level width $\Delta=0.25$ and $\epsilon_d=10$ are all fixed. \label{IVcurvelo}}
\end{figure}

\begin{figure}[t]
\includegraphics[width=0.85\columnwidth, clip]{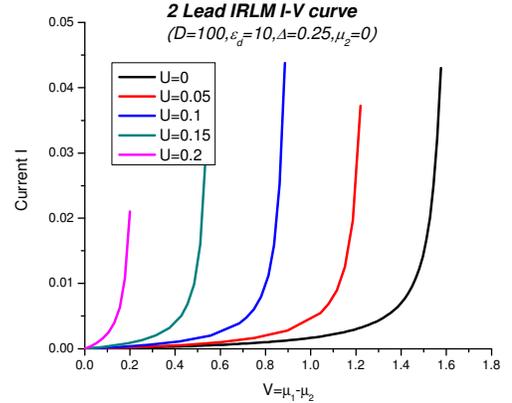}
\caption{ The current as a function of voltage for various $U$ with
$\mu_2=0$ and $\mu_1$ being increased. The bandwidth $D=100$, the
level width $\Delta=0.25$ and $\epsilon_d=10$ are all fixed.
\label{IVcurveup}}
\end{figure}

The I-V curves computed from the corrected equations are shown in
Fig. \ref{IVcurvelo},\ref{IVcurveup}.

Note that for small positive $U$ the current increases as a function
of $U$ at fixed voltage, cf \cite{BVZ}, while the current in the
Letter decreases since it corresponds to negative $U$.

 A few more comments are in order:

 $1$. The equations \ref{BAeq} are valid for $\epsilon_d > k^i_o, i=1,2$. In other cases  the scattering
 matrix has poles and zeroes and
 new terms need to be added corresponding to these "bound states" solutions, work in
 progress \cite{CMN}.\\

 $2$.  We display here the curves for small values of $U$ for which the current increases with $U$. For values
 beyond $U=2$ the current decreases with increasing $U$, cf \cite{BVZ}.

 $3$. The $I-V$ curves displayed above are not yet in their "universal form". It
is still necessary to express them in terms of RG invariant quantities such
 as the Kondo temperature $T_k$ and the anisotropy parameter while sending the
 cut-off to infinity. Some properties displayed by the curves such as the
 conductivity $G(U)=\frac{dI}{dV}|_{V=0}$, or the details of
 dependence of the current on $U$ at finite $V$, need be first expressed
 universally before they can be compared to $U$ dependence in other cut-off
 schemes used in other approaches such as perturbation theory, NRG or TD-NRG, work in progress \cite{6}.\\

 \acknowledgements  We are grateful to P. Scmitteckert and A. Zawadowski for useful and enlightening discussions.

\end{document}